\documentclass[12pt]{article}
\usepackage{epsfig}
\usepackage{amsfonts}
\usepackage{latexsym}
\usepackage{amsmath,amssymb}
\usepackage{mathrsfs}
\usepackage{hyperref}
\usepackage{setspace}
\usepackage{color}
\numberwithin{equation}{section}

\begin{document}

\begin{titlepage}
\unitlength = 1mm
\begin{flushright}
QGASLAB-14-04
\end{flushright}

\vskip 1cm
\begin{center}

 {\Large {\textsc{\textbf{Impact of quantum entanglement on spectrum of cosmological fluctuations}}}}

\vspace{1.8cm}
Sugumi Kanno

\vspace{1cm}

{\it Laboratory for Quantum Gravity \& Strings and Astrophysics, Cosmology \& Gravity Center,
Department of Mathematics \& Applied Mathematics, University of Cape Town,\\
Private Bag, Rondebosch 7701, South Africa}

\vskip 2cm

\begin{abstract}
\baselineskip=6mm
We investigate the effect of entanglement between two causally separated open
 charts in de Sitter space on the spectrum of vacuum fluctuations. 
We consider a free massive scalar field, and construct the reduced density 
matrix by tracing out the vacuum state for one of the open charts, as recently 
derived by Maldacena and Pimentel. We formulate the mean-square vacuum 
fluctuations by using the reduced density matrix and show that the scale 
invariant spectrum of massless scalar field is realized on small scales. 
On the other hand, we find that the quantum entanglement affects
the shape of the spectrum on large scales comparable to or greater than
the curvature radius.
\end{abstract}

\vspace{1.0cm}

\end{center}
\end{titlepage}

\pagestyle{plain}
\setcounter{page}{1}
\newcounter{bean}
\baselineskip18pt

\setcounter{tocdepth}{2}

\tableofcontents

\section{Introduction}

Quantum entanglement is one of the most fundamental and fascinating features of quantum mechanics. The most mysterious aspect of quantum entanglement would be to affect the outcome of local measurements instantaneously beyond the lightcone once a local measurement is performed. There are many phenomena in which quantum entanglement may play a role, including the bubble nucleation~\cite{Coleman:1980aw}. Recent studies of the bubble nucleation problem infer that observer frames will be strongly correlated to each other when they observe the nucleation frame~\cite{Garriga:2012qp, Garriga:2013pga, Frob:2014zka}.

The entanglement entropy as a suitable measure of entanglement of a quantum system has
been developed in condensed matter physics, quantum information and high energy physics. Especially, the entanglement entropy has now been established as a useful tool  in quantum field theory to characterize the nature of long range correlations. 

Even so, however, the explicit calculation of the entanglement entropy in quantum field theories had not been an easy task until Ryu and Takayanagi proposed a method of calculating the entanglement entropy of a strongly coupled quantum field theory with its gravity dual using holographic techniques~\cite{Ryu:2006bv}. Their formula has passed many consistency checks and proven to be extremely powerful~\cite{Takayanagi:2012kg}.

Following a great deal of attention paid to the success, Maldacena and Pimentel developed an explicit method to calculate the entanglement entropy in a quantum field theory in the Bunch-Davies vacuum of de Sitter space and discussed the gravitational dual of this theory and its holographic interpretation~\cite{Maldacena:2012xp}. The method is also extended to $\alpha$-vacua in~\cite{Kanno:2014lma, Iizuka:2014rua}

Based on these developments that enable us to calculate long range correlations explicitly,
it would be interesting to apply this to cosmology now. We expect that the entanglement could exist beyond the Hubble horizon because de Sitter expansion 
eventually separates off a pair of particles created within a causally connected,
Hubble horizon size region. It may be possible to investigate whether a universe entangled with our own universe exists within the multiverse framework. Such a scenario may be observable through the cosmic microwave background radiation (CMB). 

In this paper, we investigate the effect of quantum entanglement on the spectrum of vacuum fluctuations in de Sitter space by using a reduced density matrix in the open chart derived by Maldacena and Pimentel as a first step of application to cosmology. It is known that the inside of a nucleated bubble looks like an open universe~\cite{Coleman:1980aw}, so this formulation will be suitable for the multiverse framework and also for open inflation directly. The open inflation models are discussed extensively in~\cite{Sasaki:1994yt, Bucher:1994gb, Linde:1995xm, Linde:1995rv, Garriga:1997wz, Garriga:1998he}.

The paper is organized as follows. In section \ref{s2}, we review the spectrum of quantum fluctuations in the open chart. In section \ref{s3}, we review the method to derive the reduced density matrix developed by Maldacena and Pimentel with some comments relevant to 
its application to cosmology. Then we discuss the spectrum of quantum fluctuations using the reduced density matrix. We show the scale invariant spectrum is realized. In section \ref{s4}, we find the effect of the entanglement appears in the spectrum at
large wavelengths comparable to the curvature radius.
Our results are summarized and discussed in section \ref{s5}.

\section{Spectrum of quantum fluctuations in the open chart}
\label{s2}

\begin{figure}[t]
\vspace{-2.5cm}
\includegraphics[height=11cm]{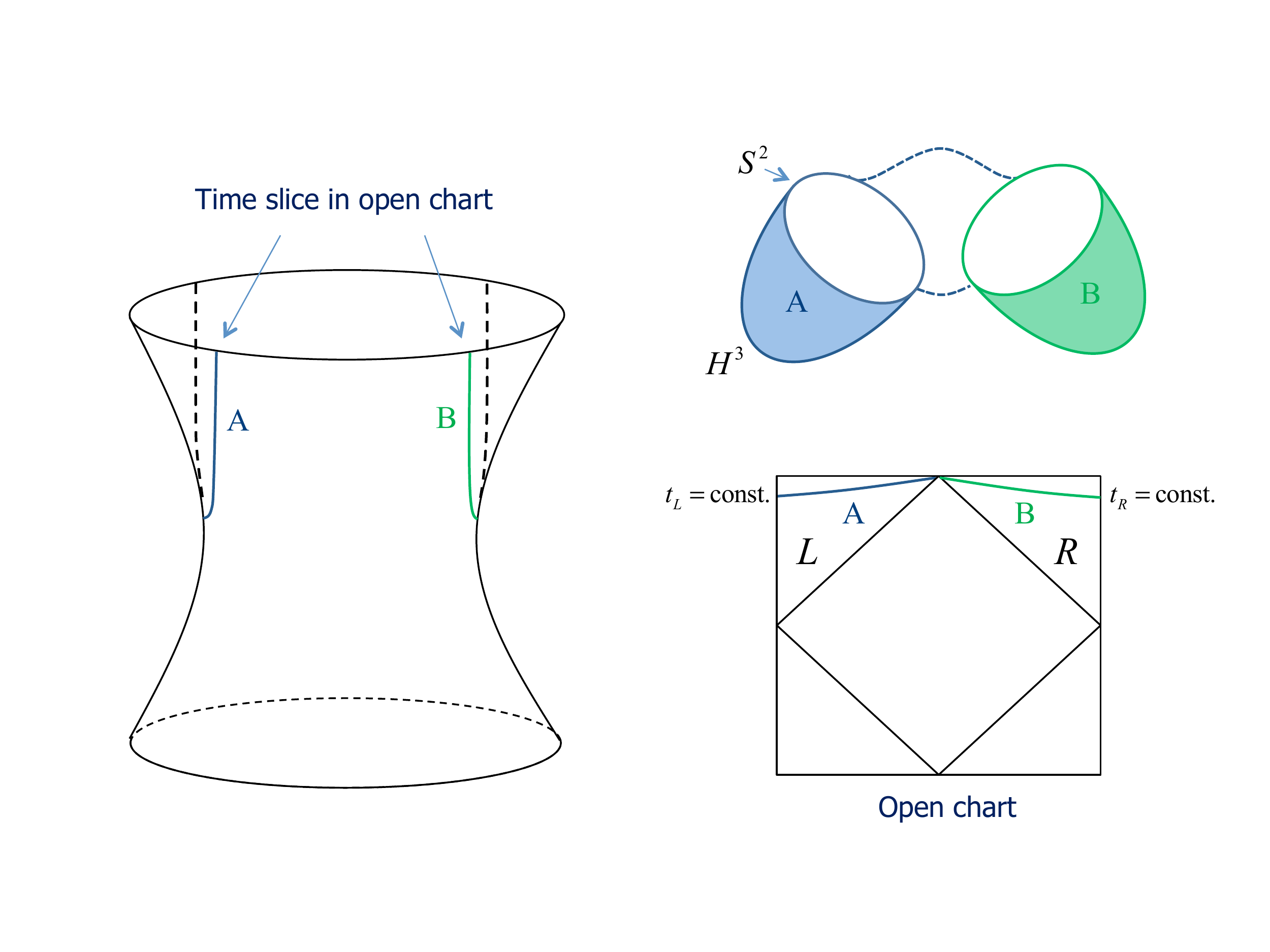}\centering
\vspace{-1cm}
\caption{De Sitter space and the Penrose diagram.}
\label{fig1}
\end{figure}

In this section, we review the spectrum of vacuum fluctuations in the open chart 
as a preparation for later sections. The Penrose diagram of the open chart is given
in Figure~\ref{fig1}, where the two time slices $A$ and $B$ represent, 
respectively, a time slice in each open chart $R$ and $L$ at sufficiently late time
in future.

\subsection{Mode functions in the open chart}
\label{s2.1}

The open de Sitter space is studied in detail in \cite{Sasaki:1994yt}. In Figure~\ref{fig1}, the de Sitter space and the Penrose diagram is depicted\footnote{The point between $L$ and $R$ regions is a part of the timelike infinity where infinite volume exits. This gives rise to the presence of the supercurvature modes for a scalar
field with sufficiently small mass~\cite{Sasaki:1994yt}. In this paper, we do not
consider the supercuvature modes.}. We consider a free scalar field of mass $m$ in de Sitter space with the action given by
\begin{eqnarray}
S=\int d^4 x \sqrt{-g} \left[\, -\frac{1}{2}\,g^{\mu\nu}\partial_\mu\phi\,\partial_\nu \phi
-\frac{m^2}{2}\phi^2\,\right]\,.
\end{eqnarray}
The metric in each region $R$ and $L$ may be obtained by analytic continuation 
from an Eulidean four-sphere metric and expressed, respectively, as
\begin{eqnarray}
ds^2_R&=&H^{-2}\left[-dt^2_R+\sinh^2t_R\left(dr^2_R+\sinh^2r_R\,d\Omega^2\right)
\right]\,,\nonumber\\
ds^2_L&=&H^{-2}\left[-dt^2_L+\sinh^2t_L\left(dr^2_L+\sinh^2r_L\,d\Omega^2\right)
\right]\,,
\end{eqnarray}
where $d\Omega^2$ is the metric on the two-sphere. Since the $R$ and $L$ regions 
are completely symmetric, if we perform the separation of variables 
\begin{eqnarray}
\phi=  \frac{H}{\sinh t}\,  \chi_p (t) \,Y_{p\ell m}(r,\Omega)\,,
\end{eqnarray} 
the equations of motion for $\chi_p$ and $Y_{p\ell m}$ in the $R$ or $L$ regions are found to be in common
\begin{eqnarray}
&&\left[\,\frac{\partial^2}{\partial t^2}+3\coth t\,\frac{\partial}{\partial t}
+\frac{1+p^2}{\sinh^2t}+\frac{m^2}{H^2}\right]\chi_p(t)=0\,,
\label{eom}\\
&&\left[\,\frac{\partial^2}{\partial r^2}+2\coth r\,\frac{\partial}{\partial r}
-\frac{1}{\sinh^2r}\,\rm\bf L^2\,\right]Y_{p\ell m}(r,\Omega)=-(1+p^2)Y_{p\ell m}(r,\Omega)\,,
\end{eqnarray}
where $(t,r)=(t_R, r_R)$ or $(t_L, r_L)$, $\rm\bf L^2$ is the Laplacian operator on the unit two-sphere, and $Y_{p\ell m}$ are harmonic functions on the 
three-dimensional hyperbolic space. 
We consider the positive frequency mode functions corresponding to the Euclidean vacuum (the Bunch-Davies vacuum \cite{Bunch:1978yq,Chernikov:1968zm,Hartle:1983ai}), because it is natural that the initial state is the de Sitter invariant vacuum. They are found explicitly in \cite{Sasaki:1994yt} and the time dependent part of $\chi_p(t)$ is given by
\begin{eqnarray}
\chi_{p,\sigma}(t)=\left\{
\begin{array}{l}
\frac{1}{2\sinh\pi p}\left(\frac{e^{\pi p}-i\sigma e^{-i\pi\nu}}{\Gamma(\nu+ip+1/2)}P_{\nu-1/2}^{ip}(\cosh t_R)
-\frac{e^{-\pi p}-i\sigma e^{-i\pi\nu}}{\Gamma(\nu-ip+1/2)}P_{\nu-1/2}^{-ip}(\cosh t_R)
\right)\,,\\
\\
\frac{\sigma}{2\sinh\pi p}\left(\frac{e^{\pi p}-i\sigma e^{-i\pi\nu}}{\Gamma(\nu+ip+1/2)}P_{\nu-1/2}^{ip}(\cosh t_L)
-\frac{e^{-\pi p}-i\sigma e^{-i\pi\nu}}{\Gamma(\nu-ip+1/2)}P_{\nu-1/2}^{-ip}(\cosh t_L)
\right)\,,\\
\end{array}
\right.
\label{chi1}
\end{eqnarray}
where $P^{\pm ip}_{\nu-\frac{1}{2}}$ are the associated Legendre functions, 
and the index $\sigma$ takes the values $\pm 1$ and distinguishes two 
independent solutions for each region. We have defined a parameter
\begin{eqnarray}
\nu=\sqrt{\frac{9}{4}-\frac{m^2}{H^2}}\,.
\end{eqnarray}
The above is a solution supported both on the $R$ and $L$ regions.
The factor $e^{-\pi p}$ in the above solutions comes from the requirement of 
analyticity in the Euclidean hemisphere which selects the Bunch-Davies vacuum. 

We expand the field in terms of the creation and anihilation operators,
\begin{eqnarray}
\hat\phi(t,r,\Omega) = \int dp \sum_{\sigma,\ell,m} 
\left[\,a_{\sigma p\ell m}\,u_{\sigma p\ell m}(t,r,\Omega)
+a_{\sigma p\ell m}^\dagger\,u^*_{\sigma p\ell m}(t,r,\Omega)\,\right]\,,
\label{phi1}
\end{eqnarray}
where $a_{\sigma p\ell m}$ satisfies $a_{\sigma p\ell m}|{\rm BD}\rangle=0$. The mode function $u_{\sigma p\ell m}(t,r,\Omega)$ representing the Bunch-Davies vacuum is
\begin{eqnarray}
u_{\sigma p\ell m} = \frac{H}{\sinh t}\,
\chi_{p,\sigma}(t)\,Y_{p\ell m} (r, \Omega)\,.
\label{mf}
\end{eqnarray}

Without loss of generality, we assume the normalization of $Y_{p\ell m}$ is such that
$Y_{p\ell m}^*=Y_{p\ell-m}$. Then Eq.~(\ref{phi1}) may be written as
\begin{eqnarray}
\hat\phi(t,r,\Omega) &=& \frac{H}{\sinh t}\int dp \sum_{\sigma,\ell,m} 
\left[\,a_{\sigma p\ell m}\,\chi_{p,\sigma}(t)
+a_{\sigma p\ell -m}^\dagger\,\chi^*_{p,\sigma}(t)\,\right]Y_{p\ell m}(r,\Omega)
\nonumber\\
&=&\int dp \sum_{\ell,m}\phi_{p\ell m}(t)Y_{p\ell m}(r,\Omega)
\,,
\label{phi1-1}
\end{eqnarray}
where we introduced a Fourier mode field operator
\begin{eqnarray}
\phi_{p\ell m}(t)\equiv\frac{H}{\sinh t}\sum_{\sigma}
\left[\,a_{\sigma p\ell m}\,\chi_{p,\sigma}(t)
+a_{\sigma p\ell -m}^\dagger\,\chi^*_{p,\sigma}(t)\,\right]\,.
\label{phi1t}
\end{eqnarray}

For convenience, we write the associated Legendre functions of the $R$ and $L$ regions in a simple form $P^{R, L}\equiv P_{\nu-1/2}^{ip}(\cosh t_{R,L})\,,\,P^{R*, L*}\equiv P_{\nu-1/2}^{-ip}(\cosh t_{R,L})$, then the two lines of Eq.~(\ref{chi1}) may be
expressed in one line as
\begin{eqnarray}
\chi^{\sigma} = N_p^{-1} \sum_{q=R,L} \left[\,
 \alpha^\sigma{}_{\!q}\,P^{q} + \beta^\sigma{}_{\!q}\,P^{q*} 
\,\right]\,,
\label{chi2}
\end{eqnarray}
where the index $p$ of $\chi_{p,\sigma}$ is omitted and we defined
\begin{eqnarray}
&&\alpha^\sigma{}_{\!R} = \frac{e^{\pi p} -i\sigma e^{-i\pi \nu}}{\Gamma (\nu+ip +\frac{1}{2})}\,,\hspace{1.1cm}
\beta^\sigma{}_{\!R} =-\frac{e^{-\pi p} -i\sigma e^{-i\pi \nu}}{\Gamma (\nu-ip +\frac{1}{2})} \,\,,\\
&&\alpha^\sigma{}_{\!L} =\sigma\,\frac{e^{\pi p} -i\sigma e^{-i\pi \nu}}{\Gamma (\nu+ip +\frac{1}{2})}  
\,,\qquad
\beta^\sigma{}_{\!L} =-\sigma\,\frac{e^{-\pi p} -i\sigma e^{-i\pi \nu}}{\Gamma (\nu-ip +\frac{1}{2})}   \,\,,
\end{eqnarray}
and $N_p$ is a normalization factor including the $1/(2\sinh\pi p)$ in Eq.~(\ref{chi1}),
given by
\begin{eqnarray}
N_{p}=\frac{4\sinh\pi p\,\sqrt{\cosh\pi p-\sigma\sin\pi\nu}}{\sqrt{\pi}\,|\Gamma(\nu+ip+1/2)|}\,.
\label{norm}
\end{eqnarray}
In the above and in what follows, it is understood that the function $P^{q}$ ($q=R$ or $L$) defined only in the $q$ region is associated with a step function which is unity in the $q$ region and which vanishes in the opposite region.
The complex conjugate of Eq.~(\ref{chi2}) which is necessary in Eq.~(\ref{phi1-1}) is
\begin{eqnarray}
\chi^{\sigma*}=N_p^{-1}\sum_{q=R,L} \left[\,
\beta^{\sigma*}{}_{\!\!\!\!q}\,P^q + \alpha^{\sigma*}{}_{\!\!\!\!q}\,P^{q*} 
\,\right]\,.
\label{cmpxchi2}
\end{eqnarray}
If we introduce a $4\times 4$ matrix
\begin{eqnarray}
\chi^I=\left(
\begin{array}{l}
\chi^\sigma\\
\chi^{\sigma*}
\end{array}\right)\,,\qquad
M^I{}_J=\left(
\begin{array}{ll}
\alpha^\sigma{}_{\!q} &~ \beta^\sigma{}_{\!q} \vspace{3mm}\\
\beta^{\sigma*}{}_{\!\!\!q} &~ \alpha^{\sigma*}{}_{\!\!\!q} \\
\end{array}\right)\,,\qquad
P^J=\left(
\begin{array}{l}
P^q\\
P^{q*}
\end{array}\right)\,,
\label{M}
\end{eqnarray}
where $\sigma=\pm 1$, $q=(R, L)$ and the summation is understood 
when a pair of the same indices appear upstairs and downstairs,
 then both of Eqs.~(\ref{chi2}) and (\ref{cmpxchi2}) can be accommodated into 
the simple matrix form,
\begin{eqnarray}
\chi^I=N_p^{-1}\,M^I{}_J\,P^J\,,
\label{chiI}
\end{eqnarray}
where the capital indices $(I,J)$ run from 1 to 4. The Fourier mode field operator Eq.~(\ref{phi1t}) is then expanded as
\begin{eqnarray}
\phi_{p\ell m}(t)\equiv\phi(t)=\frac{H}{\sinh t}\,a_{I}\,\chi^{I}
=\frac{1}{N_p}\frac{H}{\sinh t}\,a_{I}\,M^{I}{}_{J}\,P^{J}\,,\qquad\quad
a_{I}=\left(\,a_\sigma\,,\,a_\sigma^\dagger\,\right)\,,
\label{phit}
\end{eqnarray}
where $t$ stands for a time slice given by $t=t_R$ in the $R$ region and $t=t_L$ in the $L$ region, and in the following the indices $p,\ell, m$ of $\phi_{p\ell m}(t)$ are omitted for simplicity unless there may be any confusion.

\subsection{Spectrum in the $L$ region}
\label{s2.2}

Let us calculate the power spectrum, say, in the $L$ region because the $R$ and $L$
 regions are completely symmetric. As it is unnecessary to consider the $R$ region, 
the mode function $\chi^I$ given by Eq.~(\ref{chiI}) may be restricted to
the $L$ region. This reduces the $4\times4$ matrix $M^I{}_J$ to 
the $4\times 2$ matrix $M^{I}{}_{\cal J}$, where the calligraphic indice ${\cal J}$
runs from 1 to 2, and the solution on the $L$ region in Eq.~(\ref{chi1}) 
is expressed as
\begin{eqnarray}
\chi^{I}=N_p^{-1}\,M^{I}{}_{\cal J}\,P^{\cal J}\,,
\label{chi3}
\end{eqnarray}
where 
\begin{eqnarray}
\chi^{I}=\left(
\begin{array}{l}
\chi^\sigma\\
\chi^{\sigma*}
\end{array}\right)\,,\quad
M^{I}{}_{\cal J}=\left(
\begin{array}{ll}
\alpha^\sigma{}_{\!L} &~ \beta^\sigma{}_{\!L} \vspace{3mm}\\
\beta^{\sigma*}{}_{\!\!\!\!L} &~ \alpha^{\sigma*}{}_{\!\!\!\!L} \\
\end{array}\right)\,,\quad
P^{\cal J}=\left(
\begin{array}{l}
P^L\\
P^{L*}
\end{array}\right)\,.
\label{M2}
\end{eqnarray}
The Fourier mode field operator in Eq.~(\ref{phit}) is now expressed as
\begin{eqnarray}
\phi(t_L)=\frac{H}{\sinh t_L}\,a_{I}\,\chi^{I}
=\frac{1}{N_p}\frac{H}{\sinh t_L}\,a_{I}\,M^{I}{}_{\cal J}\,P^{\cal J}\,,\qquad\quad
a_{I}=\left(\,a_\sigma\,,\,a_\sigma^\dagger\,\right)\,,
\label{phi2}
\end{eqnarray}
where we used Eq.~(\ref{chi3}).

If we focus on a single mode with indices $p,\ell, m$, 
the mean-square vacuum fluctuation
is then computed by using Eq.~(\ref{phit}) as
\begin{eqnarray}
\langle{\rm BD}|\,\phi_{p\ell m}(t_L)\,\phi_{p'\ell'm'}^\dagger(t_L)\,|{\rm BD}\rangle
&=&
\frac{H^2}{\sinh^2t_L}
\langle{\rm BD}|\,a_{I}\chi^{I}\,\left(a_{J}\chi^{J}\right)^\dagger\,|{\rm BD}\rangle
\nonumber\\
&=&
\frac{H^2}{\sinh^2t_L}\sum_{\sigma=\pm1}|\chi^\sigma|^2\,
\delta(p-p')\delta_{\ell\ell'}\delta_{mm'}
\nonumber\\
&\equiv&
S(p,t_L)\,\delta(p-p')\delta_{\ell\ell'}\delta_{mm'}\,.
\label{ev1}
\end{eqnarray}
Then the normalized spectrum per unit logarithmic interval of $p$
is given by
\begin{eqnarray}
{\cal P}(p,t_L)=\frac{p^3}{2\pi^2}S(p,t_L)
=\frac{p^3}{2\pi^2}\frac{H^2}{\sinh^2t_L}\sum_{\sigma=\pm1}|\chi^\sigma|^2\,,
\label{def:P1}
\end{eqnarray}
where $\chi^\sigma=\chi_{p,\sigma}(t_L)$.

\subsection{Vacuum fluctuations after horizon exit}
\label{s2.3}

In this subsection, we make sure that the amplitude of the vacuum fluctuation is frozen out at the epoch of horizon exit. The time dependence of $\chi^\sigma$ in Eq.~(\ref{chi3}) comes from the associated Legendre functions. So if we write
\begin{eqnarray}
\chi^{I}=\left(
\begin{array}{l}
\chi^\sigma\\
\chi^{\sigma*}
\end{array}\right)
=\left(
\begin{array}{l}
A^\sigma{}_{\!L}\,P^L+B^\sigma{}_{\!L}\, P^{L*}\\
B^{\sigma*}{}_{\!\!\!\!L}\,P^L+A^{\sigma*}{}_{\!\!\!\!L}\,P^{L*}
\end{array}\right)\,,\qquad
A^\sigma{}_{\!L}\equiv\frac{\alpha^\sigma{}_{\!L}}{N_p}\,,\qquad
B^\sigma{}_{\!L}\equiv\frac{\beta^\sigma{}_{\!L}}{N_p}\,,
\label{AB1}
\end{eqnarray}
then $|\chi^\sigma|^2$ in the normalized spectrum in Eq.~(\ref{def:P1}) is
expressed as
\begin{eqnarray}
|\chi^\sigma|^2=A^\sigma{}_{\!L} B^{\sigma*}{}_{\!\!\!\!L}\left(P^L\right)^2
+\left(|A^\sigma{}_{\!L}|^2+|B^\sigma{}_{\!L}|^2\right)P^LP^{L*}
+A^{\sigma*}{}_{\!\!\!\!L}B^\sigma{}_{\!L}\left(P^{L*}\right)^2\,.
\label{chisquared1}
\end{eqnarray}
Since the Legendre functions on superhorizon scale behave as
\begin{eqnarray}
P_{\nu-1/2}^{\pm ip}(\cosh t)\xrightarrow{t\gg 1}
\frac{2^{\nu-1/2}\Gamma(\nu)}{\sqrt{\pi}\,\Gamma(\nu\mp ip+1/2)}
\left(\cosh t\right)^{\nu-\frac{1}{2}}\,,
\label{legendre2}
\end{eqnarray}
we find the time dependence of each combination of the associated Legendre functions in Eq.~(\ref{chisquared1}) is the same. Then $\sum_{\sigma=\pm1}|\chi^\sigma|^2$ is given by
\begin{eqnarray}
&&\hspace{-7mm}
\sum_{\sigma=\pm1}|\chi^\sigma|^2\xrightarrow{t_L\gg 1}
\frac{2^{2\nu-1}\Gamma(\nu)^2}{\pi}\sum_{\sigma=\pm1}
\left[\frac{A^\sigma{}_{\!L} B^{\sigma*}{}_{\!\!\!\!L}}{\Gamma(\nu-ip+1/2)^2}
+\frac{|A^\sigma{}_{\!L}|^2+|B^\sigma{}_{\!L}|^2}{|\Gamma(\nu+ip+1/2)|^2}
+\frac{A^{\sigma*}{}_{\!\!\!\!L}B^\sigma{}_{\!L}}{\Gamma(\nu+ip+1/2)^2}
\right]
\nonumber\\
&&\hspace{2.3cm}
\times\left(\cosh t_L\right)^{2\nu-1}\nonumber\\
&&\hspace{2.3cm}\equiv Z\left(\cosh t_L\right)^{2\nu-1}\,,
\label{z1}
\end{eqnarray}
where we defined the time independent part of $\sum_{\sigma=\pm1}|\chi^\sigma|^2$ by $Z$. 

Plugging Eq.~(\ref{z1}) into Eq.~(\ref{def:P1}), we find the spectrum approaches
\begin{eqnarray}
{\cal P}(p,t_L)=\frac{H^2}{\sinh^2t_L}\sum_{\sigma=\pm1}|\chi^\sigma|^2\,\frac{p^3}{2\pi^2}\,
\xrightarrow{t_L\gg 1}
H^2Z\,\frac{p^3}{2\pi^2}\,\frac{\left(\cosh t_L\right)^{2\nu-1}}{\sinh^2t_L}\,.
\label{time1}
\end{eqnarray}
For massless case $\nu=3/2$, we find the time dependent part becomes
\begin{eqnarray}
\frac{\left(\cosh t_L\right)^{2\nu-1}}{\sinh^2t_L}\xrightarrow{t_L\gg 1}1\,.
\end{eqnarray}
Thus, the vacuum fluctuation gets frozen out after horizon exit.

\subsection{Wavenumber dependence}
\label{s2.4}

\begin{figure}[t]
\vspace{-2.5cm}
\includegraphics[height=10cm]{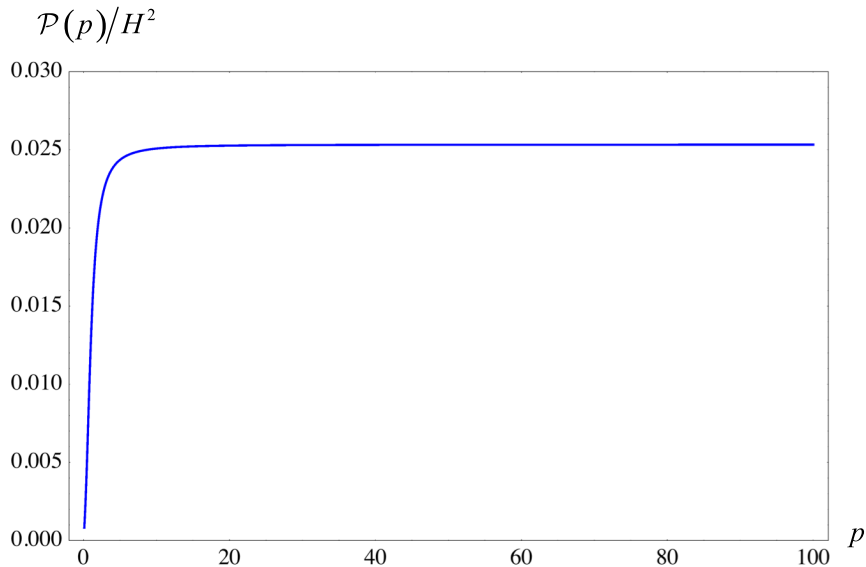}\centering
\vspace{-1cm}
\caption{The spectrum of vacuum fluctuations for a massless scalar field 
as a function of $p$.}
\label{fig2}
\end{figure}

We also need to check the behavior of the spectrum at short wavelengths ($p\gg 1$).
The spectrum should be the same as the case of a specially flat universe. For the 
massless scalar field ($\nu=3/2$), the dominant term in $Z$ of Eq.~(\ref{z1}) 
for large $p$ is
\begin{eqnarray}
\sum_{\sigma=\pm1}\frac{|A^\sigma{}_{\!L}|^2}{|\Gamma(\nu+ip+1/2)|^2}
=\sum_{\sigma=\pm1}\frac{|\alpha^\sigma{}_{\!L}|^2/|N_p|^2}{|\Gamma(\nu+ip+1/2)|^2}
\xrightarrow{p\gg 1}
\frac{2\pi^2e^{-2\pi p}}{p^5|\Gamma(i p)|^4}\,.
\label{dominant1}
\end{eqnarray}
Then the time independent part of $Z$ at short wavelengths ($p\gg 1$) becomes
\begin{eqnarray}
Z\xrightarrow{p\gg 1}
\frac{2^2\Gamma(3/2)^2}{\pi}\frac{2\pi^2e^{-2\pi p}}{p^5|\Gamma(i p)|^4}
\sim\frac{1}{2p^3}\,.
\end{eqnarray}
Thus the spectrum of the massless scalar field after horizon exit is evaluated
as
\begin{eqnarray}
{\cal P}(p)=H^2Z\,\frac{p^3}{2\pi^2}\,\xrightarrow{p\gg 1}
\left(\frac{H}{2\pi}\right)^2\,.
\label{sg1}
\end{eqnarray}
This is the well-known result for vacuum fluctuations after horizon exit. 
The spectrum as a function of $p$ is plotted in Figure~\ref{fig2}.

\section{Spectrum of quantum fluctuations using the reduced density matrix}
\label{s3}

In previous section, we discussed the spectrum of vacuum fluctuations by using 
the field operator expanded in terms of the mode functions defined for
the Bunch-Davies vacuum. We next want to explore the spectrum of vacuum 
fluctuations by using the reduced density matrix derived by Maldacena and 
Pimentel in \cite{Maldacena:2012xp}. In this section, we first review the 
formalism to obtain the reduced density matrix and then discuss the 
spectrum of vacuum fluctuations.

\subsection{A review on the reduced density matrix}
\label{s3.1}

In order to derive the reduced density matrix in the region $L$, we need to 
trace over the degrees of freedom of the region $R$. To make this procedure 
practically possible, the reduced density matrix has to be diagonalized. 
The following is a review on the derivation of the reduced density matrix 
given in \cite{Maldacena:2012xp}.

As we see in Eq.~(\ref{phit}), the mode function $\chi^I$ corresponding to the 
Bunch-Davies vacuum is a linear combination of the Legendre functions 
$P^{R,L}=P^{ip}_{\nu-1/2}(\cosh t_{R,L})$, which are proportional to the 
positive frequency modes in the past in each $R$ or $L$ region. 
Setting $\tau=\cosh t$, the Legendre functions are known to be
normalized as
\begin{eqnarray}
\left(\tau^2-1\right)\left(\frac{dP^{q*}}{d\tau}P^q-\frac{dP^q}{d\tau}P^{q*}\right)
=\frac{2ip}{|\Gamma(1+ip)|^2}\,,
\end{eqnarray}
where $q=R$ or $L$.
Thus, the properly normalized positive frequency functions
in each $R$ or $L$ region are given by
\begin{eqnarray}
\varphi^q=N_b^{-1}P^q\,,\qquad N_b=\frac{\sqrt{2p}}{|\Gamma(1+ip)|}\,.
\label{norm1}
\end{eqnarray}


Let us introduce $\varphi^I$ as
\begin{eqnarray}
\varphi^I=\left(
\begin{array}{l}
\varphi^{q}\\
\varphi^{q*}
\end{array}\right)\,.
\end{eqnarray}
Then this procedure of changing the mode functions from $\chi^I$ to $\varphi^I$ is 
a Bogoliubov transformation. The Bogoliubov coefficients are then expressed 
in terms of $\alpha$ and $\beta$ in the matrix $M$ in Eq.~(\ref{M}). 
So let us introduce new creation and anihilation operators $b_I$ defined 
such that $b_R|R\rangle=0$ and $b_L|L\rangle=0$.
Now, with the proper normalization in Eq.~(\ref{norm1}), the original field 
operator in Eq.~(\ref{phit}) is expanded in terms of the new operators $b_J$ as
\begin{eqnarray}
\phi(t)=\frac{H}{\sinh t}\,a_I\,\chi^I
=\frac{H}{\sinh t}\,b_J\,\varphi^J\,,\qquad
b_J=\left(\,b_{q}\,,\,b_{q}^\dagger\,\right)\,,
\label{phi3}
\end{eqnarray}
where again the capital indices $(I,J)$ run from 1 to 4,
$q=(R,L)$, and the repeated indices are summed over. 
By comparing Eqs.~(\ref{phit}) and (\ref{phi3}), the relation between the 
operators $a_I$ and $b_I$ has to be
\begin{eqnarray}
a_J=N_p\,N_b^{-1}\,b_I\left(M^{-1}\right)^I{}_J\,,\qquad
\left(M^{-1}\right)^I{}_J=\left(
\begin{array}{ll}
\xi^q{}_{\sigma} &~ \delta^q{}_{\sigma} \vspace{3mm}\\
\delta^{q*}{}_{\!\!\sigma} &~ \xi^{q*}{}_{\!\!\sigma} \\
\end{array}\right)\,,\qquad
\left\{
\begin{array}{l}
\xi=
\left(\alpha-\beta\,\alpha^{*\,-1}\beta^*\right)^{-1}\,,\vspace{3mm}\\
\delta=-\alpha^{-1}\beta\,\xi^*\,.
\end{array}
\right.
\label{ab}
\end{eqnarray}
Note that $M^{-1}$ is not normalized but its determinant is 
$\left(|\xi|^2-|\delta|^2\right)^{-1/2}=\left(|\alpha|^2-|\beta|^2\right)^{1/2}$,
which corresponds to $N_p\,N_b^{-1}$ in Eq.~(\ref{ab}).  
Thus, the Bunch-Davies vacuum can be regarded as a Bogoliubov transformation 
of the $R,L$-vacua as
\begin{eqnarray}
|{\rm BD}\rangle\propto\exp\left(\frac{1}{2}\sum_{i,j=R,L}m_{ij}\,b_i^\dagger\, b_j^\dagger\right) |R\rangle|L\rangle\,,
\label{bogoliubov1}
\end{eqnarray}
where $m_{ij}$ is a symmetric matrix and the operators $b_i$ satisfy the commutation relation $[b_i,b_j^\dagger]=\delta_{ij}$. Note that the normalization of the Bogoliubov transformation is omitted here for simplicity because it is unnecessary 
for the derivation of the reduced density matrix, which will be given 
in Eq.~(\ref{dm}).

Here we mention a subtlety associated with the fact that the $p$-mode spectrum
is continuous. Since the commutation relation between the $p$-mode annihilation
operator and the $p'$-mode creation operator is proportional to $\delta(p-p')$,
that between the same spectral index $p=p'$ would diverge. To avoid
this divergence, we discretize the $p$-mode spectrum with a width
$\Delta p$ and take the limit $\Delta p\to0$ only at the end of computation.
This means, for example, we rescale the operators $b_i$ as
$b_i\to b_i=\sqrt{\Delta p}\,b_i^{\rm cont}$, where the index `cont'
denotes the operator with the original, continuous spectrum, and the operators
$b_i$ and $b_j$ appearing in Eq.~(\ref{bogoliubov1}) should be understood
as the rescaled ones.

 The condition $a_\sigma|{\rm BD}\rangle=0$ determines $m_{ij}$:
\begin{eqnarray}
m_{ij}=-\delta_{i\sigma}^*\left(\xi^{-1}\right)_{\sigma j}
=e^{i\theta}\frac{\sqrt{2}\,e^{-p\pi}}{\sqrt{\cosh 2\pi p+\cos 2\pi\nu}}
\left(
\begin{array}{cc}
\cos \pi\nu & i\sinh p\pi \vspace{1mm}\\
i\sinh p\pi & \cos \pi\nu \\
\end{array}
\right)\,,
\label{mij}
\end{eqnarray}
where $e^{i\theta}$ contains all unimportant phase factors for $\nu^2>0$. 

When the state is written in the form of Eq.~(\ref{bogoliubov1}), 
it is still difficult to trace over the $R$ degrees of freedom because the
density matrix $\rho=|{\rm BD}\rangle\langle{\rm BD}|$ is not diagonal in
the $|R\rangle|L\rangle$ basis.
Thus, we perform another Bogoliubov transformation further
by introducing new operators $c_R$ and $c_L$
\begin{eqnarray}
c_R = u\,b_R + v\,b_R^\dagger \,,\qquad\quad
c_L = \bar{u}\,b_L + \bar{v}\,b_L^\dagger\,,
\label{bc0}
\end{eqnarray}
to obtain the relation
\begin{eqnarray}
|{\rm BD}\rangle = N_{\gamma_p}^{-1}
\exp\left(\gamma_p\,c_R^\dagger\,c_L^\dagger\,\right)|R'\rangle|L'\rangle\,.
\label{bogoliubov2}
\end{eqnarray}
Here, the normalizaion $|u|^2-|v|^2=1$ and $|\bar{u}|^2-|\bar{v}|^2=1$ are assumed
so that the new operators satisfy the commutation relation 
$[c_i,c_j^\dagger]=\delta_{ij}$. 
The normalization factor $N_{\gamma_p}$ is given by
\begin{eqnarray}
N_{\gamma_p}^2
=\left|\exp\left(\gamma_p\,c_R^\dagger\,c_L^\dagger\,\right)|R'\rangle|L'\rangle
\right|^2
=\frac{1}{1-|\gamma_p|^2}\,,
\label{norm2}
\end{eqnarray}
where $|\gamma_p|<1$ is imposed.
Eq.~(\ref{bc0}) is a linear transformation between $c_q$ and $b_q$, 
so this Bogoliubov transformation does not mix $R$ and $L$ Hilbert spaces, but
the basis vacuum changes from $|R\rangle|L\rangle$ to 
$|R^\prime\rangle|L^\prime\rangle$. 

For later convenience, we introduce a $4\times4$ matrix form of Eq.~(\ref{bc0}),
\begin{eqnarray}
c_J=b_I\,G^I{}_J\,,\qquad
G^I{}_J=\left(
\begin{array}{ll}
U_{q} &~ V_{q}^* \\
V_{q} &~ U_{q}^* \\
\end{array}\right)\,,\qquad
c_J=(c_{q}\,,c_{q}^\dagger)\,,
\label{bc}
\end{eqnarray}
where $U_{q}\equiv{\rm diag}(u,\bar{u})$, $V_{q}\equiv{\rm diag}(v,\bar{v})$. 

The consistency conditions for Eq.~(\ref{bogoliubov2}) are
\begin{eqnarray}
c_R\,|{\rm BD}\rangle= \gamma_p\,c_L^\dagger\,|{\rm BD}\rangle \,,\qquad 
c_L\,|{\rm BD}\rangle = \gamma_p\,c_R^\dagger\,|{\rm BD}\rangle\,.
\label{consistency}
\end{eqnarray}
If we write $m_{RR} = m_{LL}\equiv \omega$ and $m_{LR}=m_{RL}\equiv \zeta$
in Eq.~(\ref{mij}), we see that $\omega$ is real and $\zeta$ is pure imaginary 
for positive $\nu^2$. Then inserting Eqs.~(\ref{bogoliubov1}) and (\ref{bc0}) 
into Eq.~(\ref{consistency}), we find a system of four homogeneous equations
\begin{eqnarray}
&&\omega\,u + v -\gamma_p\,\zeta\,\bar{v}^* =0 \ , \qquad 
\zeta\,u - \gamma_p\,\bar{u}^* - \gamma_p\,\omega\,\bar{v}^* =0\,,
\label{system1}\\
&&\omega\,\bar{u} + \bar{v} -\gamma_p\,\zeta\,v^* =0 \ , \qquad
\zeta\,\bar{u} - \gamma_p\,u^* - \gamma_p\,\omega\,v^* =0\,,
\label{system2}
\end{eqnarray}
where $\omega^*=\omega$ and $\zeta^*=-\zeta$.
We see that setting $v^* =\bar{v}$ and $u^* =\bar{u}$ is possible 
if $\gamma_p$ is pure imaginary $\gamma_p^*=-\gamma_p$. 
This is always possible by adjusting the phase of $c_q$.
Then Eq.~(\ref{system2}) becomes  identical with Eq.~(\ref{system1}) and 
the system is reduced  to that of two homogeneous equations. 
We look for such $\gamma_p$, keeping the normalization condition 
$|u|^2-|v|^2=1$ satisfied. 

In order to have a non-trivial solution in the system of equations (\ref{system1}),
 $\gamma_p$ must be
\begin{eqnarray}
\gamma_p=\frac{1}{2\zeta}\left[-\omega^2+\zeta^2+1-\sqrt{\left(\omega^2-\zeta^2-1\right)^2-4\zeta^2}\,\right]\,,
\label{gammap}
\end{eqnarray}
where we took a minus sign in front of the square root term to 
satisfy $|\gamma_p|<1$. Note that $\gamma_p$ is pure imaginary.
Putting the $\omega$ and $\zeta$ defined in Eq.~(\ref{mij}) 
into Eq.~(\ref{gammap}), we get 
\begin{eqnarray}
\gamma_p = i\frac{\sqrt{2}}{\sqrt{\cosh 2\pi p + \cos 2\pi \nu}
 + \sqrt{\cosh 2\pi p + \cos 2\pi \nu +2 }}\,.
\label{gammap2}
\end{eqnarray}

Now we have the density matrix which enables us
to trace over the $R$ degrees of freedom easily.
By using  Eqs.~(\ref{bogoliubov2}) and (\ref{norm2}), the reduced density 
matrix is then found to be diagonalized as
\begin{eqnarray}
\rho_L ={\rm Tr}_{R}\,|{\rm BD}\rangle\langle{\rm BD}| 
=\left(1-|\gamma_p|^2\,\right)\sum_{n=0}^\infty 
|\gamma_p |^{2n}\,|n;p\ell m\rangle\langle n;p\ell m|\,,
\label{dm} 
\end{eqnarray}
where we defined $|n;p\ell m\rangle=1/\sqrt{n!}\,(c_L^\dagger)^n\,|L'\rangle$. 
Note that this density matrix is for each mode labeled by $p,\ell, m$.

\subsection{Spectrum of vacuum fluctuations}
\label{s3.2}

Now we calculate the mean-square vacuum fluctuations by using the reduced density 
matrix obtained in Eq.~(\ref{dm}). Since we traced out the degrees of freedom of 
the $R$ region, the corresponding field operator has to be defined only in the 
$L$ region. Also the field operator should act only on $|L'\rangle$ conforming with 
the reduced density matrix. Then, the Fourier mode field operator should
be defined as 
\begin{eqnarray}
\phi_{L p\ell m}(t_L)
\equiv\phi_{L}(t_L)
&=&\frac{1}{N_b}\frac{H}{\sinh t_L}\,b_{\cal J}\,P^{\cal J}
=\frac{1}{N_b}\frac{H}{\sinh t_L}\,c_{\cal I}\left(G^{-1}\right)^{\cal I}{}_{\cal J}\,P^{\cal J}\nonumber\\
&\equiv&\frac{H}{\sinh t_L}\,c_{\cal I}\,\psi^{\cal I}\,,\qquad
\psi^{\cal I}=\left(
\begin{array}{l}
\psi^{L}\\
\psi^{L*}
\end{array}\right)\,,
\label{phi4}
\end{eqnarray}
where the calligraphic indices $({\cal I, J})$ run from 1 to 2, and 
the matrix is reduced to a $2\times2$ form:
\begin{eqnarray}
\left(G^{-1}\right)^{\cal I}{}_{\cal J}
=\left(
\begin{array}{cc}
\bar{u}^* & -\bar{v}^* \\
-\bar{v} & \bar{u} \\
\end{array}\right)
\,,\qquad
|\bar{u}|^2-|\bar{v}|^2=1\,.
\end{eqnarray}
Note that we transformed the field operator expanded by $b_J$ in Eq.~(\ref{phi3})
into $c_I$ by using Eq.~(\ref{bc})\footnote{The Fourier mode field operator $\phi_{L}$ is rescaled in accordance with the discretization of the $p$-mode spectrum as explained below Eq.~(\ref{bogoliubov1}).}.

Focusing on a single mode with indices $p,\ell,m$,
the mean-square vacuum fluctuations are then calculated as
\begin{eqnarray}
{\rm Tr}_L\rho_L\,\phi_{L}\,\phi_{L}^\dagger
&=&\left(1-|\gamma_p|^2\,\right)\sum_{n=0}^\infty |\gamma_p |^{2n}
\langle n;p\ell m|\,\phi_{L}\,\phi_{L}^\dagger\,|n;p\ell m\rangle\,,
\label{ev2}
\end{eqnarray}
where $\phi_L=\phi_{Lp\ell m}(t_L)$.

\subsection{Spectrum in the $L$ region}
\label{s3.3}

We calculate the power spectrum in the $L$ region. The mode function given in 
Eq.~(\ref{phi4}) is written as
\begin{eqnarray}
\psi^{\cal I}=N_b^{-1}\left(G^{-1}\right)^{\cal I}{}_{\cal J}\,P^{\cal J}
=\left(
\begin{array}{l}
{\cal A}_{L}P^L+{\cal B}_LP^{L*}\\
{\cal B}^*_{L}P^L+{\cal A}^*_LP^{L*}
\end{array}\right)\,,
\label{psi}
\end{eqnarray}
where we defined
\begin{eqnarray}
{\cal A}_{L}=\frac{\bar{u}^*}{N_b}\,,\qquad
{\cal B}_{L}=-\frac{\bar{v}^*}{N_b}\,.
\end{eqnarray}
Here, $\bar{u}$, $\bar{v}$ are obtained by solving Eq.~(\ref{system1}) with the 
solution Eq.~(\ref{gammap2}) and imposing the normalization condition 
$|\bar{u}|^2-|\bar{v}|^2=1$:
\begin{eqnarray}
\bar{u}=\frac{1-\gamma_p\zeta}{\sqrt{|1-\gamma_p\zeta|^2-|\omega|^2}}
\,,\qquad
\bar{v}=\frac{\omega}{\sqrt{|1-\gamma_p\zeta|^2-|\omega|^2}}\,.
\end{eqnarray}
Here, $\omega=m_{RR} = m_{LL}$ and $\zeta=m_{LR}=m_{RL}$ in Eq.~(\ref{mij}). 
Using these we can compute the expectation value for each
$n$-particle state in Eq.~(\ref{ev2}), 
\begin{eqnarray}
\langle n;p\ell m|\,\phi_{L}\,\phi_{L}^\dagger\,|n;p\ell m\rangle
&=&\frac{1}{n!}\,\langle L'|(c_L)^n\,{\phi}_{L}\,{\phi}_{L}^\dagger\,(c_L^\dagger)^n|L'\rangle
\nonumber\\
&=&
\frac{H^2}{\sinh^2t_L}\,\frac{1}{n!}\,
\langle L'|(c_L)^n\, c_{\cal I}\,\psi^{\cal I}\,\left(c_{\cal J}\,\psi^{\cal J}\right)^\dagger\,(c_L^\dagger)^n|L'\rangle
\nonumber\\
&=&\frac{H^2}{\sinh^2t_L}|\psi^L|^2\left(2n+1\right)\,,
\label{ev2-1}
\end{eqnarray}
where from the second line to the third line we used
\begin{eqnarray}
&&\langle L'|(c_L)^n\,
c_{\cal I}\,\psi^{\cal I}\,\left(c_{\cal J}\,\psi^{\cal J}\right)^\dagger\,(c_L^\dagger)^n|L'\rangle\nonumber\\
&&\qquad=|\psi^L|^2\,n!\,\langle n;p\ell m|n;p\ell m\rangle
+2|\psi^L|^2\,n^2(n-1)!\,\langle n-1;p\ell m|n-1;p\ell m\rangle
\nonumber\\
&&\qquad=(2n+1)\,n!\,|\psi^L|^2\,.
\end{eqnarray}
Putting Eq.~(\ref{ev2-1}) into Eq.~(\ref{ev2}), the mean-square vacuum fluctuations
for each mode is given by
\begin{eqnarray}
{\rm Tr}_L\rho_L\,\phi_{L}\,\phi_{L}^\dagger&=&
\frac{H^2}{\sinh^2t_L}|\psi^L|^2\left(1-|\gamma_{p}|^2\,\right)
\sum_{n=0}^\infty |\gamma_{p}|^{2n}(2n+1)
\nonumber\\
&=&\frac{H^2}{\sinh^2t_L}\,|\psi^L|^2\,
\frac{1+|\gamma_{p}|^2}{1-|\gamma_{p}|^2}\,.
\label{ev3}
\end{eqnarray}
Note that by comparing with Eq.~(\ref{ev1}),
we see the main effect of tracing out the degrees of freedom of 
the region $R$ seems to come in with
the form of $(1+|\gamma_p|^2)/(1-|\gamma_p|^2)$.
 Since $|\gamma_p|\rightarrow 1$ as $p\rightarrow 0$ (for $m=0$), this extra 
term enhances the mean-square vacuum fluctuations at long wavelengths ($p\ll 1$). 
Note also that the mode functions are changed from
\begin{eqnarray}
\chi^{I}=N_p^{-1}\,M^{I}{}_{\cal J}\,P^{\cal J}\,\longrightarrow\,
\psi^{\cal I}=N_b^{-1}\left(G^{-1}\right)^{\cal I}{}_{\cal J}\,P^{\cal J}\,,
\label{chipsi}
\end{eqnarray}
due to the change from the Bunch-Davies vacuum to $L'$-vacuum 
which diagonalized the reduced density matrix as obtained in Eq.~(\ref{dm}).

\subsection{Vacuum fluctuations after horizon exit}
\label{s3.4}

As we did in subsection \ref{s2.3}, let us check the time dependence of $\psi^{\cal I}$ in Eq.~(\ref{phi5}). From Eq.~(\ref{psi}), the $|\psi^{\cal I}|^2$ in 
the mean-square vacuum fluctuations in Eq.~(\ref{ev3}) is expressed as
\begin{eqnarray}
|\psi^L|^2={\cal A}_L{\cal B}_L^*\left(P^L\right)^2
+\left(|{\cal A}_L|^2+|{\cal B}_L|^2\right)P^LP^{L*}
+{\cal A}_L^*{\cal B}_L\left(P^{L*}\right)^2\,.
\label{psi2}
\end{eqnarray}
Since the Legendre functions on superhorizon scale is the same as in Eq.~(\ref{legendre2}), we find the time dependence of this case is the same as Eq.~(\ref{z1}) and 
expressed as
\begin{eqnarray}
&&|\psi^L|^2\xrightarrow{t_L\gg 1}
\frac{2^{2\nu-1}\Gamma(\nu)^2}{\pi}
\left[\frac{{\cal A}_L{\cal B}_L^*}{\Gamma(\nu-ip+1/2)^2}
+\frac{|{\cal A}_L|^2+|{\cal B}_L|^2}{|\Gamma(\nu+ip+1/2)|^2}
+\frac{{\cal A}_L^*{\cal B}_L}{\Gamma(\nu+ip+1/2)^2}
\right]
\nonumber\\
&&\hspace{2cm}\times\left(\cosh t_L\right)^{2\nu-1}\nonumber\\
&&\hspace{2cm}\equiv {\cal Z}\left(\cosh t_L\right)^{2\nu-1}\,,
\label{z2}
\end{eqnarray}
where we defined ${\cal Z}$ corresponding to $Z$ in Eq.~(\ref{z1}) for comparison.
The spectrum of the vacuum fluctuations corresponding to Eq.~(\ref{def:P1})
 is now given by
\begin{eqnarray}
{\cal P}(p, t_L)=\frac{p^3}{2\pi^2}\,
{\rm Tr}_L\rho_L\,\phi_{L}\,\phi_{L}\,.
\end{eqnarray}
Plugging Eqs.~(\ref{ev3}) and (\ref{z2}) into above, we find the spectrum is found to be
\begin{eqnarray}
{\cal P}(p, t_L)=\frac{H^2}{\sinh^2t_L}\,|\psi^L|^2\,\frac{1+|\gamma_p|^2}{1-|\gamma_p|^2}
\,\frac{p^3}{2\pi^2}\,
\xrightarrow{t_L\gg 1}
H^2{\cal Z}\,\frac{1+|\gamma_p|^2}{1-|\gamma_p|^2}
\,\frac{p^3}{2\pi^2}\,
\frac{(\cosh t_L)^{2\nu-1}}{\sinh^2t_L}\,.
\label{time2}
\end{eqnarray}
We see the time dependent part is completely identical with Eq.~(\ref{time1}). Thus, for massless case ($\nu=3/2$), the vacuum fluctuation gets frozen after horizon exit in the case of the spectrum using the reduced density matrix as well.

\subsection{Wavenumber dependence}
\label{s3.5}

\begin{figure}[t]
\vspace{-2.5cm}
\includegraphics[height=10cm]{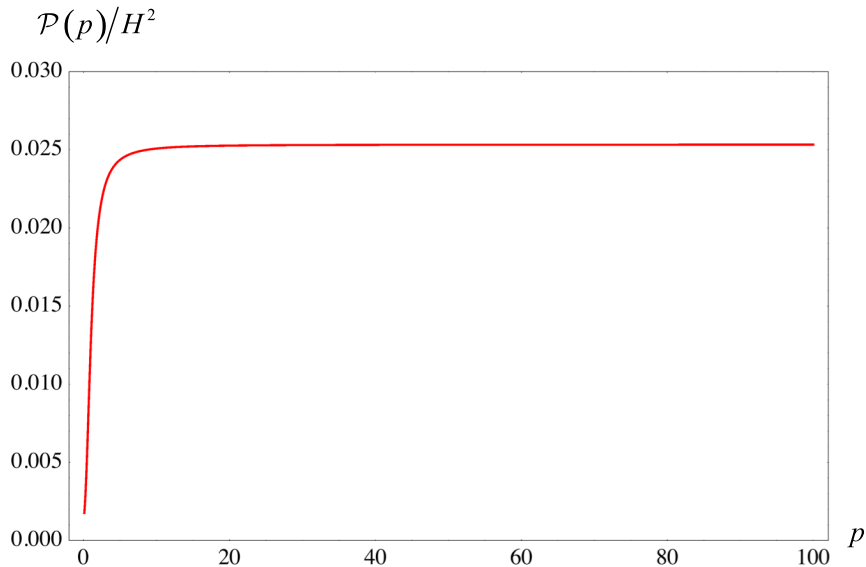}\centering
\vspace{-1cm}
\caption{The spectrum as a function of $p$.}
\label{fig3}
\end{figure}

Next, let us see the behavior of the power spectrum at short wavelengths ($p\gg 1$) 
as we did in subsection \ref{s2.4}. For a massless scalar field ($\nu=3/2$), 
the dominant term in ${\cal Z}$ of Eq.~(\ref{z2}) for large $p$ is found to be
\begin{eqnarray}
\frac{|{\cal A}_L|^2}{|\Gamma(\nu+ip+1/2)|^2}
=\frac{|\bar{u}|^2/|N_b|^2}{|\Gamma(\nu+ip+1/2)|^2}
\xrightarrow{p\gg 1}
\frac{\pi e^{-\pi p}}{2p^4|\Gamma(i p)|^2}\,.
\label{dominant2}
\end{eqnarray}
The corresponding case of the Bunch-Davies vacuum is Eq.~(\ref{dominant1}). 
We see the $\alpha_L^\sigma/N_p$ is simply replaced by $\bar{u}/N_b$,
and find that the behavior of ${\cal Z}$ is the same as that of $Z$,
\begin{eqnarray}
{\cal Z}\xrightarrow{p\gg 1}
\frac{2^2\Gamma(3/2)^2}{\pi}\frac{\pi e^{-\pi p}}{2p^4|\Gamma(i p)|^2}
\sim\frac{1}{2p^3}\,.
\end{eqnarray}
Thus the spectrum of a massless scalar field after horizon exit is evaluated as
\begin{eqnarray}
{\cal P}(p)=H^2{\cal Z}\,\frac{1+|\gamma_p|^2}{1-|\gamma_p|^2}\left(\frac{p^3}{2\pi^2}\right)
\xrightarrow{p\gg 1}\left(\frac{H}{2\pi}\right)^2\,.
\label{sg2}
\end{eqnarray}
We find that the scale invariant spectrum is realized for $p\gg1$
even after tracing out the degrees of freedom of the region $R$. 
We plot the spectrum of the massless scalar field in Eq.~(\ref{sg2}) as a function of $p$ in Figure~\ref{fig3}.
It is almost identical to Figure~\ref{fig2}. 

\section{Possible observable signatures}
\label{s4}

So far, we discussed two different spectra given in terms of
$\chi^I$ and $\psi^{\cal I}$. The former correspond to the mode
functions for the {\it entangled}\, Bunch-Davies vacuum,
and the latter to those that diagonalize the reduced density
matrix obtained from the Bunch-Davies vacuum by tracing out 
the degrees of freedom of the $R$ region. 
The corresponding states are given by the density matrix 
$\rho_{\rm BD}=|{\rm BD}\rangle\langle{\rm BD}|$
and the reduced density matrix $\rho_L={\rm Tr}_{R}\,\rho_{\rm BD}$,
respectively. 

The difference between these two states is the observer's point of 
view when taking the mean-square vacuum fluctuations. For $\rho_{\rm BD}$,
we are supposed to know the entangled state in the region $R$. 
On the other hand, for $\rho_L$, we are supposed to be completely 
ignorant about the state in the $R$ region.
In order to distinguish them, we call $\rho_{\rm BD}$ the ``entangled state'', 
and $\rho_L$ the ``mixed state (obtained from the entangled state)'',
or just denote them by their relevant mode functions $\chi^I$ and $\psi^{\cal I}$
 in the following.

\subsection{Non-entangled state}
\label{s4.1}

In the subsections~\ref{s2.4} and \ref{s3.5}, we showed that both of the entangled state and the mixed state realize the scale invariant spectrum for a massless 
scalar field at short wavelengths ($p\gg 1$). 
Now we wonder what would happen to the spectrum if we assume the $L$ region 
were the whole universe and use the mode function $\varphi^{\cal I}$ 
which is defined only in the $L$ region. We call it 
``non-entangled state", and denote it by $\varphi^{\cal I}$.

The relevant mode function for the non-entangled case
should be given by Eq.~(\ref{norm1}). That is
\begin{eqnarray}
\varphi^{\cal I}=\left(
\begin{array}{l}
\varphi^{L}\\
\varphi^{L*}
\end{array}\right)
=\frac{1}{N_b}
\left(
\begin{array}{l}
P^L\\
P^{L*}
\end{array}\right)\,.
\end{eqnarray}
The spectrum is then defined by
\begin{eqnarray}
{\cal P}(p, t_L)=\frac{p^3}{2\pi^2}\,\langle L|\,\varphi^L\varphi^L\,|L\rangle\,,
\end{eqnarray}
where the Fourier mode field operator is
\begin{eqnarray}
\varphi^L(t_L)&=&\frac{H}{\sinh t}\,
b_{\cal I}\,\varphi^{\cal I}(t_L)\,,
\label{phi5}
\end{eqnarray}
and we find the mean-square vacuum fluctuations for each mode
\begin{eqnarray}
\langle L|\,\varphi^L\varphi^L\,|L\rangle
=\frac{H^2}{\sinh^2t_L}|\varphi^L|^2\,\delta(p-p')\delta_{\ell\ell'}\delta_{mm'}\,.
\end{eqnarray}
The $|\varphi^L|^2$ becomes a simple form
\begin{eqnarray}
&&|\varphi^L|^2=\frac{1}{2p}\left|\Gamma(1+ip)\right|^2P^LP^{*L}
\nonumber\\
&&\hspace{1cm}\xrightarrow{t_L\gg 1}
\frac{2^{2\nu-1}\Gamma(\nu)^2}{\pi}
\frac{|\Gamma(1+ip)|^2}{|\Gamma(\nu+ip+1/2)|^2}\frac{1}{2p}
\left(\cosh t_L\right)^{2\nu-1}\equiv \tilde{\cal Z}\left(\cosh t_L\right)^{2\nu-1}\,.
\label{time3}
\end{eqnarray}
Here, time independent part of $\tilde{\cal Z}$ at short wavelengths ($p\gg 1$)
is evaluated as
\begin{eqnarray}
\tilde{\cal Z}\xrightarrow{p\gg 1}
\frac{2^2\Gamma(3/2)^2}{\pi}\frac{|\Gamma(1+ip)|^2}{|\Gamma(2+i p)|^2}\frac{1}{2p}
\sim\frac{1}{2p^3}\,.
\end{eqnarray}
The power spectrum becomes
\begin{eqnarray}
{\cal P}(p)=H^2\tilde{\cal Z}\,\frac{p^3}{2\pi^2}\,
\xrightarrow{p\gg 1}\left(\frac{H}{2\pi}\right)^2\,.
\end{eqnarray}
Thus, for $p\gg1$, we get the flat spectrum of vacuum fluctuations for a
massless scalar field even for the non-entangled state. 
This means that we cannot distinguish which state is valid for our universe 
as far as we focus on the short wavelength region of the spectrum.
So let us examine the behavior at long wavelengths after the horizon exit.

\subsection{Spectra at long wavelengths}
\label{s4.2}

\begin{figure}[t]
\begin{center}
\vspace{-2cm}
\begin{minipage}{8.1cm}
\hspace{-1.2cm}
\includegraphics[height=7cm]{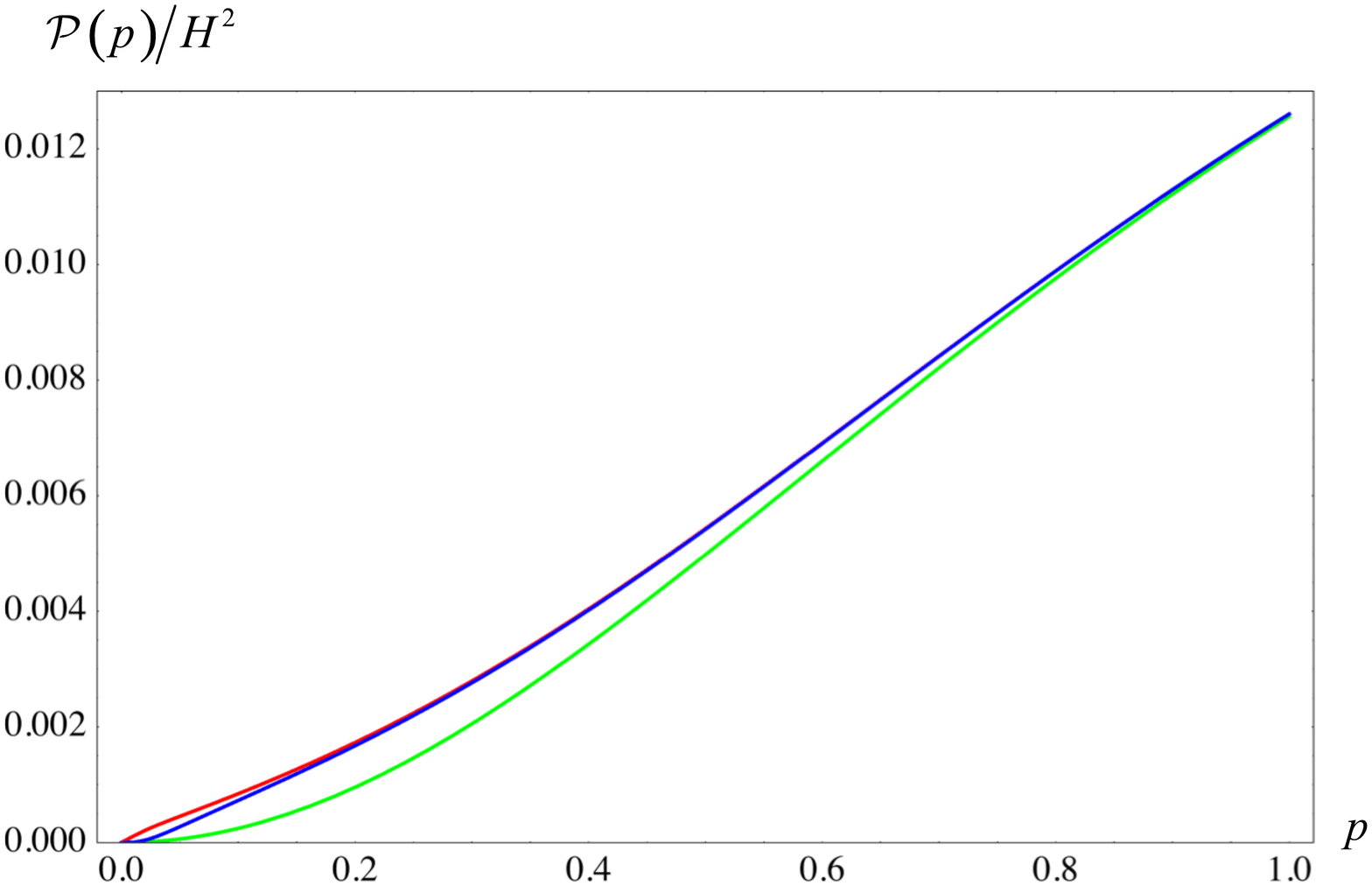}
\vspace{-0.7cm}
\end{minipage}
\begin{minipage}{8.1cm}
\hspace{-0.5cm}
\includegraphics[height=7cm]{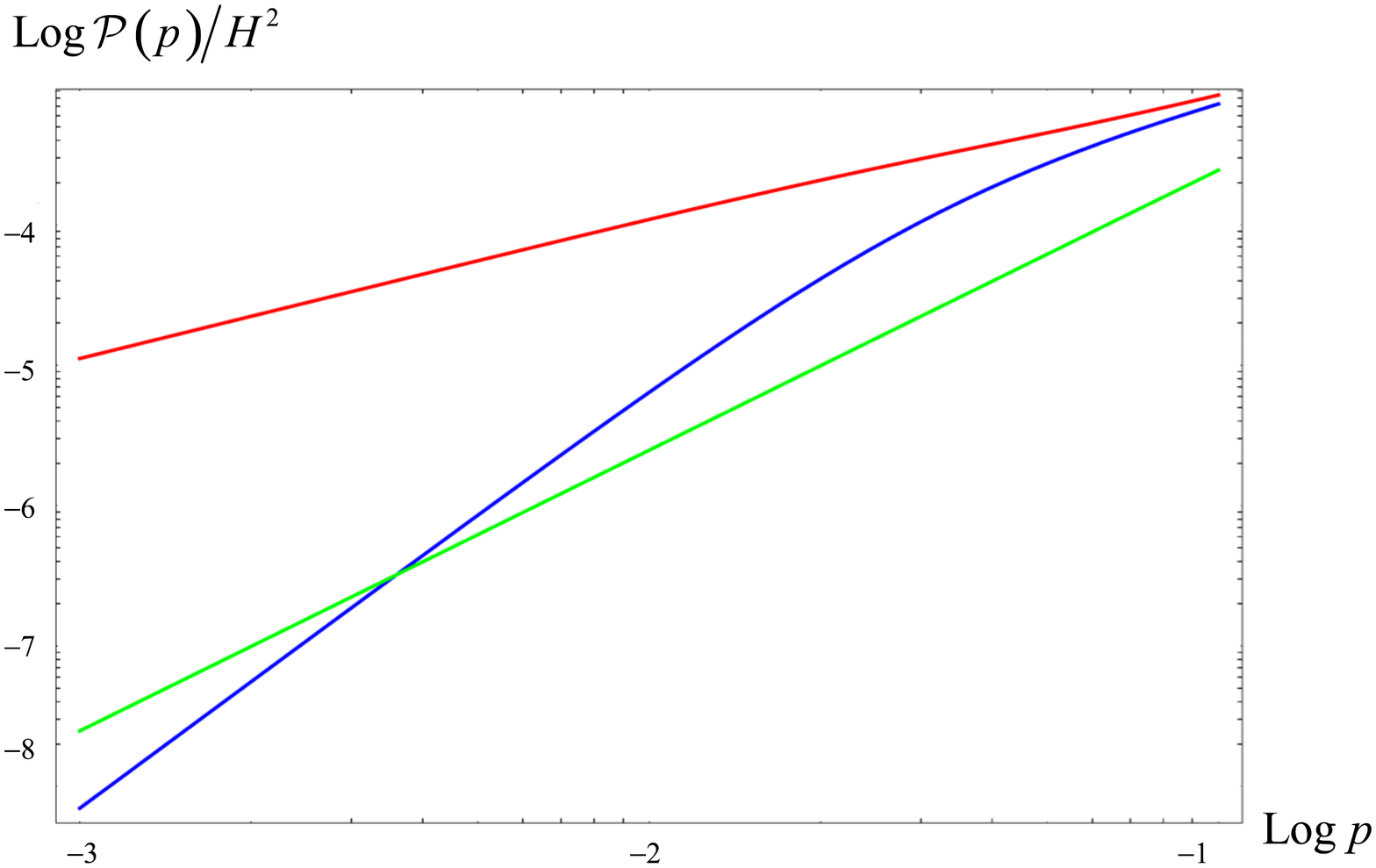}
\vspace{-1.3cm}
\end{minipage}
\caption{The spectrum at long wavelengths (Left) and the logarithmic plot at
very long wavelengths (Right). 
The red line is for  $\psi^{\cal I}$, the blue line is for $\chi^I$, 
and the green is for $\varphi^{\cal I}$. 
}
\label{fig4}
\end{center}
\end{figure}

In order to see the behavior at long wavelengths, we take the
limit $p\ll1$ and expand the expressions for the spectra in $p$.
For $\varphi^{\cal I}$ and $\chi^I$, we expand $\tilde{\cal Z}$ 
in Eq.~(\ref{time3}) and $Z$ in Eq.~(\ref{z1}), respectively.
Then we find the $p$ dependence at leading order is given by
\begin{eqnarray}
{\cal P}(p)\xrightarrow[p\ll 1]{t_L\gg 1}\left\{
\begin{array}{l}
2^{2\nu-1}\Gamma(\nu)^2\pi^{-1}|\Gamma(\nu+1/2)|^{-2}
\left(\frac{H}{2\pi}\right)^2p^2\hspace{1.4cm}
{\rm for}\quad\varphi^{\cal I}\,,\\\\
2^{2\nu-1}\Gamma(\nu)^2|\Gamma(\nu+1/2)|^{-2}
\left(\frac{H}{2\pi}\right)^2p^3
\,\hspace{2cm} {\rm for}\quad\chi^I\,.
\end{array}
\right.
\label{smallp1}
\end{eqnarray} 
In the case of $\psi^{\cal I}$, we expand $\cal Z$ in Eq.~(\ref{z2})
to obtain
\begin{eqnarray}
&&\hspace{-0.5cm}{\cal P}(p)\xrightarrow[p\ll 1]{t_L\gg 1}
2^{2\nu-1}\Gamma(\nu)^2\,\sqrt{2}\,\pi^{-2}\,|\Gamma(\nu+1/2)|^{-2}\,|\cos\pi\nu|
\left(\sqrt{1+\cos2\pi\nu}+\sqrt{3+\cos2\pi\nu}\right)
\nonumber\\
&&\hspace{2cm}\times\frac{(\sqrt{1+\cos2\pi\nu}+\sqrt{3+\cos2\pi\nu})^2+2}{(\sqrt{1+\cos2\pi\nu}+\sqrt{3+\cos2\pi\nu})^2-2}
\left(\frac{H}{2\pi}\right)^2p\,\hspace{1.5cm} {\rm for}\quad
\psi^{\cal I}\,.
\label{smallp2}
\end{eqnarray}
We are interested in the case of small mass ($|\nu-3/2|\ll 1$).
Taking this limit, we see that the spectrum for 
$\psi^{\cal I}$ decreases 
most slowly as $p\rightarrow 0$.
The spectra at long wavelengths are plotted in Figure \ref{fig4}, where we take 
$m^2=H^2/10$. The spectra for the entangled state ($\chi^I$) and 
the mixed state ($\psi^{\cal I}$) are the blue line and the red line, 
respectively. The green line is the spectrum for the non-entangled 
state ($\varphi^{\cal I}$). The non-entangled state seems to give the
most suppressed spectrum at $p\ll1$. But if we use the logarithmic plots 
(right panel)
we find the result is consistent with the analysis of 
the $p$ dependence in Eqs.~(\ref{smallp1}) and (\ref{smallp2}).

Around the curvature scale of the open universe $(p\sim1)$, 
we find that the green 
line starts to deviate from other two lines as $p\rightarrow 0$. 
This point tells us if the regions $R$ and $L$ are entangled or not. 
Around $5$ times of the curvature scale of the open universe $(p\sim0.2)$, the red 
and blue lines start to bifurcate. This tells us which state among the 
entangled state and the mixed state is appropriate for our universe.


\section{Summary and discussion}
\label{s5}

We investigated the spectrum of vacuum fluctuations with quantum entanglement 
in de Sitter space by using the reduced density matrix obtained by 
Maldacena and Pimentel in~\cite{Maldacena:2012xp}. We formulated the 
mean-square vacuum fluctuations by introducing the two open charts, $L$ and $R$,
which are entangled. We found that the effect of tracing out the degrees 
of freedom of the $R$ region comes in to the mean-square of vacuum fluctuations 
and enhances it on scales larger than the curvature radius.
On the other hand, on small scales, the spectrum of vacuum fluctuations 
of a massless scalar field is found to be scale invariant, and completely
indistinguishable from the pure Bunch-Davies (hence highly entangled) case.

We also considered a fictitious case in which we assume the $L$ region were
the whole universe. In this case, with respect to a natural vacuum state
defined in the $L$ region, the scale invariant spectrum is also realized 
on small scales. Thus it turned out that we cannot distinguish which state 
is valid for our universe 
among the entangled state ($\chi^I$), the mixed state from the 
entangled state ($\psi^{\cal I}$) and non-entangled state ($\varphi^{\cal I}$),
as long as we focus on the spectrum on small scales. 

On the other hand, the spectrum for each of these three cases is found
to be different from each other on large scales comparable to or greater than
the curvature radius. 

We found that the spectra for the entangled and mixed states tend 
to be enhanced compared to the non-entangled state on scales comparable
to the curvature radius.
If we go further to scales much larger than the curvature radius, 
there appears a difference between the entangled state and mixed state.
The spectrum for the mixed state is enhanced substantially relative
to the entangled state. Since we can use these differences to
distinguish those three states, they seem to be relevant (and hopefully
observationally testable) in the context of open inflation.

Now let us consider why the difference appeared in the spectra between 
the entangled and mixed states. In general, if we consider two subsystems 
$A$ and $B$ that form a state $|\Psi\rangle$, 
the expectation value of a physical quantity ${\cal O}$ is 
expressed by using the density matrix $\rho=|\Psi\rangle\langle\Psi|$ as
\begin{eqnarray}
{\rm Tr}\rho\,{\cal O}=\sum_{\alpha,\beta}\langle\alpha,\beta|\Psi\rangle\langle\Psi
|{\cal O}|\alpha,\beta\rangle
=\langle\Psi|{\cal O}|\Psi\rangle\,.
\label{general:ev2}
\end{eqnarray}
If ${\cal O}$ depends only on $A$, that is, ${\cal O}_A$ :
\begin{eqnarray}
\langle\alpha',\beta'|{\cal O}|\alpha,\beta\rangle=\langle\alpha|{\cal O}_A|\alpha'\rangle
\delta_{\beta\beta'}\,,
\end{eqnarray}
then Eq.~(\ref{general:ev2}) becomes
\begin{eqnarray}
{\rm Tr}\rho\,{\cal O}_A&=&\sum_{\alpha,\beta}\sum_{\alpha',\beta'}
\langle\alpha,\beta|\Psi\rangle
\langle\Psi|\alpha',\beta'\rangle\langle\alpha',\beta'|
{\cal O}_A|\alpha,\beta\rangle
=\sum_{\alpha,\beta}\sum_{\alpha'}
\langle\alpha,\beta|\Psi\rangle\langle\Psi|\alpha',\beta\rangle
\langle\alpha'|{\cal O}_A|\alpha\rangle\nonumber\\
&=&{\rm Tr}\rho_A\,{\cal O}_A\,,
\end{eqnarray}
where
\begin{eqnarray}
\rho_A\equiv{\rm Tr}_B\rho=\sum_{\alpha,\alpha'}|\alpha\rangle\left(\sum_\beta\langle\alpha,\beta|\Psi\rangle\langle\Psi|\alpha',\beta\rangle\right)\langle\alpha'|\,.
\end{eqnarray}
Thus, the expectation value ${\rm Tr}\rho\,{\cal O}$ is equal to
 ${\rm Tr}\rho_A\,{\cal O}_A$ if ${\cal O}$ does not depend on $B$.
In other words, for an operator with the property ${\cal O}={\cal O}_A$, 
the expectation value should agree with each other.

Based on this general argument, we may understand the reason why the spectra for
the entangled state and the mixed state from the entangled state become 
different on large scale $(p\rightarrow 0)$.
We see that the creation and anihilation operators $a_J$ include 
the operators on both sides, $b_J=(b_q,b_q^\dag)$ ($q=R,L$), 
as in Eq.~(\ref{ab}). 
This means that the field operator $\phi(t)$ naturally involves 
operators on both sides even if the time slice $t$ is restricted
to the $L$ region.
On the other hand, after tracing out the degrees of freedom of
the $R$ region, the mixed state $(\psi^{\cal I})$ is described
by the creation and anihilation operators defined only in the $L$ region.
That is, the field operator $\phi_L(t)$ is defined only in the $L$ region.
Thus apparently $\phi(t)\neq\phi_L(t)$, implying that the above general
argument does not hold because ${\cal O}\neq{\cal O}_A$.

Finally, we mention that it would be easy to extend our analysis to 
the case of gravitons. Another direction is to consider interactions.
The formalism we developed in this paper is at tree level. 
It would be interesting to consider the loop corrections and see
how they affect the result. 
They might enhance the effect of the entanglement on small scales
because of the coupling between long and short wavelengths~\cite{Tanaka:2013caa}.

\section*{Acknowledgments}
I would like to thank Jiro Soda and Misao Sasaki for fruitful discussions, and for helpful suggestions and comments. I would also like to thank Alex Vilenkin and Jaume Garriga for careful reading of the draft, and for very useful suggestions and comments. This work was supported in part by funding from the University Research Council 
of the University of Cape Town.

\end{document}